\documentclass[a4paper,11pt]{article}
\usepackage{amsmath,amssymb,amsfonts}
\usepackage{algorithmic}
\usepackage{graphicx}
\usepackage{textcomp}
\usepackage{caption}
\usepackage{subfig}
\usepackage[margin=1in]{geometry}
\usepackage{authblk}
\usepackage[dvipsnames]{xcolor}

\usepackage[colorlinks=true,
            citecolor=blue,
            linkcolor=ForestGreen,
            urlcolor=blue]{hyperref}

\usepackage[numbers,sort&compress]{natbib}

\title{\bfseries Idling error suppression through gate scheduling}

\author[1,2]{Hoiki Madison Liu\thanks{Corresponding author:
\texttt{hoiki.liu@fujitsu.com}}}
\author[1,2]{Kazunori Maruyama}
\author[1,2]{Hirotaka Oshima}
\author[1,2]{Shintaro Sato}

\affil[1]{Quantum Laboratory, Fujitsu Research, Fujitsu Limited, 4-1-1 Kawasaki, Kanagawa, Japan}

\affil[2]{RIKEN Center for Quantum Computing, 2-1 Hirosawa, Wako, Saitama, Japan}

\begin{document}
\maketitle
\begingroup
\renewcommand{\thefootnote}{}
\footnotetext{\small
This work has been submitted to the IEEE for possible publication.
Copyright may be transferred without notice, after which this version
may no longer be accessible.}
\endgroup
\setcounter{footnote}{0}
\begin{abstract}
Achieving high-precision quantum computation requires effective suppression of idling errors that occur when qubits remain inactive during waiting periods within a quantum circuit. Conventional mitigation techniques, such as dynamical decoupling, suppress decoherence by periodically refreshing quantum states through the insertion of additional control gates. In this paper, we propose an alternative approach that suppresses idling errors through quantum circuit scheduling without introducing any additional gate operations. By appropriately adjusting the execution timing of quantum gates with scheduling flexibility, we demonstrate through both numerical simulations and hardware experiments that the overall computational accuracy can be significantly influenced and, in many cases, improved. In addition, we analytically derive the density-matrix evolution under idling noise and provide a theoretical framework that explains the observed behavior.
\end{abstract}


\section{Introduction}
\label{sec:introduction}
\par
One of the central challenges in the realization of practical quantum computers is decoherence, whereby the quantum states of qubits degrade due to unavoidable interactions with the external environment. Qubit states are inherently fragile, and interactions with environmental factors—such as thermal fluctuations, electromagnetic noise, and residual quantum couplings—lead to a gradual loss of coherence over time. Such decoherence occurs when a qubit is not actively manipulated, i.e., when it remains in an idling state. These effects are commonly referred to as idling errors \cite{Krantz:2019jkw, Terhal:2013vbm} and are observed across multiple quantum computing platforms, including superconducting and trapped-ion hardware \cite{Chen:2021num,Jurcevic:2020uhm,PhysRevLett.121.220502}.
\par
Idling errors primarily consist of two types: amplitude damping and dephasing. Amplitude damping corresponds to the spontaneous relaxation of a qubit from its excited state to the ground state, while dephasing arises from stochastic fluctuations that randomize the relative phase of the quantum state. In addition, frequency drift (detuning) can accumulate coherent Z-phase. And there are also correlated and hardware-specific mechanisms such as residual ZZ interactions between qubits, crosstalk from control pulses, and leakage into non-computational states. Altogether, these idling-induced error processes can significantly degrade the fidelity of quantum operations and thereby limit the overall accuracy of quantum computation.
\par
On the other hand, although quantum error correction codes such as surface codes have attracted considerable attention as a promising approach to fault-tolerant quantum computation \cite{Bravyi:1998sy,Dennis:2001nw,Freedman:1998sw,Fowler:2012hwn}, superconducting quantum devices often suffer from limitations in performance—such as gate fidelity and qubit connectivity—or in hardware topology and defects \cite{Kjaergaard:2019lmy, Bilmes:2019rum,Lisenfeld:2019pap}, which makes the direct implementation of quantum error correction codes challenging. To address these constraints, techniques such as temporarily swapping qubit positions using SWAP gates are commonly employed \cite{Reichardt:2018kqi,Zhou:2024wsl,Wei:2026icn,Nagayama:2017ifj}. However, this approach introduces circuit segments that cannot be fully parallelized. Consequently, certain qubits remain in an idling state for extended periods, leading to the accumulation of substantial idling errors. Therefore, suppressing idling errors is critically important for achieving high-fidelity implementations of complex quantum error-correction circuits.
\par
One well-established technique for mitigating the effects of idling errors is dynamical decoupling \cite{Bylander:2011zcm,Khodjasteh:2005ttl,Khodjasteh:2007smb,Paz-Silva:2012wvy,Souza:2011idd,Viola:1998gg,Das:2021sig}. This approach suppresses decoherence by repeatedly applying sequences of single-qubit gate operations that effectively average out environmental noise. Such pulse sequences implement an effective identity operation and therefore preserve the intended computational outcome. However, dynamical decoupling relies on the application of high-precision control pulses, such as $\pi$ pulses or XY pulse sequences. In the presence of pulse-amplitude errors, timing inaccuracies, or other control imperfections, these operations may induce unintended rotations and consequently introduce additional errors. Therefore, although dynamical decoupling can be highly effective on quantum devices with sufficiently high gate fidelity, its practical applicability becomes limited in regimes where control errors are non-negligible.
\par
In this paper, we propose an alternative approach that suppresses idling errors through quantum circuit scheduling without introducing any additional gate operations.In particular, we focus on the Hadamard (H) gate, which is widely used in quantum error-correction protocols and often possesses scheduling flexibility. By appropriately adjusting the execution timing of such gates, we demonstrate through both numerical simulations and hardware experiments on single- and two-qubit idling circuits that the overall computational accuracy can be significantly influenced and, in many cases, improved. Next, we analytically derive the density-matrix evolution under noise models that include not only amplitude damping and dephasing, but also coherent frequency drift, and evaluate the trace distance in order to provide a theoretical explanation for the observed behavior. Furthermore, we investigate the practically important scenario in which the initial state prior to the H gate is unknown. In this setting, we model the unknown initial state as a Haar-random pure state and evaluate the Haar-averaged trace distance under the proposed scheduling strategies.

\section{Idling error}
\subsection{Amplitude damping}
\par
Amplitude damping describes the irreversible energy relaxation process in which a qubit decays from its excited state $|1\rangle$ to the ground state $|0\rangle$ due to coupling with the environment.
\par
The amplitude damping process can be modeled as a completely positive trace-preserving (CPTP) quantum channel $\mathcal{E}_{\mathrm{AD}}$ acting on a single-qubit density matrix $\rho$. In the Kraus operator formalism, the channel is expressed as
\begin{align}
\mathcal{E}_{\mathrm{AD}}(\rho)=\sum_{k=0,1} E_k \rho E_k^{\dagger},
\end{align}
where the Kraus operators $E_0$ and $E_1$ are given by
\begin{align}
E_0=\left(\begin{array}{cc}
1 & 0 \\
0 & \sqrt{1-\gamma}
\end{array}\right), \quad E_1=\left(\begin{array}{cc}
0 & \sqrt{\gamma} \\
0 & 0
\end{array}\right) .\label{ampdamp}
\end{align}
Here, $\gamma \in [0,1]$ denotes the probability of energy decay during the relevant time interval. For an idling period of duration $t$, the decay probability is related to the relaxation time $T_1$ as $\gamma = 1 - e^{-t/T_1}$.
\par
Applying the amplitude damping channel to a general single-qubit density matrix
\begin{align}
\rho =
\begin{pmatrix}
\rho_{00} & \rho_{01} \\
\rho_{10} & \rho_{11}
\end{pmatrix}.
\end{align}
yields
\begin{align}
    \mathcal{E}_{\mathrm{AD}}(\rho) =
\begin{pmatrix}
\rho_{00} + \gamma \rho_{11} & \sqrt{1 - \gamma}\,\rho_{01} \\
\sqrt{1 - \gamma}\,\rho_{10} & (1 - \gamma)\rho_{11}
\end{pmatrix}.
\end{align}
\par
This expression highlights two key features of amplitude damping. First, population in the excited state $|1\rangle$ decays exponentially into the ground state $|0\rangle$, leading to an increase in $\rho_{00}$ and a decrease in $\rho_{11}$. Second, the off-diagonal coherence terms $\rho_{10}$ and $\rho_{01}$ are attenuated by a factor of $\sqrt{1 - \gamma}$, indicating that amplitude damping also contributes to decoherence. The ground state $|0\rangle$ is a fixed point of the amplitude damping channel, remaining unaffected by the noise. This property explains why idling in the $|0\rangle$ state does not introduce amplitude damping errors, whereas idling in states with nonzero excited-state population leads to fidelity loss.

\subsection{Dephasing}
\par
Dephasing, also referred to as phase damping, describes the loss of quantum coherence without energy exchange between the qubit and its environment. Physically, dephasing arises from fluctuations in the qubit transition frequency induced by environmental noise or imperfections in the control system, such as flux or charge noise in superconducting qubits. These fluctuations generate random phase accumulations during idle periods, causing the ensemble-averaged coherence of the quantum state to decay over time. Unlike amplitude damping, dephasing does not alter the population of the computational basis states $|0\rangle$ and $|1\rangle$, but instead randomizes their relative phase. This process is commonly characterized by the dephasing time $T_2$ (or, more precisely, the pure dephasing time $T_\phi$).
\par
The Kraus operators are given by
\begin{align}
M_0=\sqrt{1-\lambda}I,\quad M_1 =\sqrt{\lambda}Z.\label{dephasing}
\end{align}
Here, $I$ denotes the identity operator, $Z$ is the Pauli-Z operator, and $\lambda \in [0,1]$ denotes the probability of dephasing over the relevant time interval. For a Markovian process, the dephasing probability is related to the coherence time $T_2$ as $\lambda=\frac{1-e^{-t/T_2}}{2}$. The state evolution is then
\begin{align}
\mathcal{E}_{\mathrm{DP}}(\rho)&=\sum_{k=0,1} M_k \rho M_k^{\dagger}=
\begin{pmatrix}
\rho_{00} & (1-2\lambda)\rho_{01} \\
(1-2\lambda)\rho_{10} &\rho_{11}
\end{pmatrix}.
\end{align}
This result illustrates that dephasing preserves the populations of the computational basis states while exponentially suppressing the off-diagonal coherence terms. Consequently, states aligned with the $Z$ basis, such as $|0\rangle$ and $|1\rangle$, are unaffected by dephasing, whereas superposition states, including $|+\rangle$ and $|-\rangle$, are particularly vulnerable.
\par
\par
Using (\ref{ampdamp}) and (\ref{dephasing}), the combined channel of amplitude damping and dephasing can be described by a set of three Kraus operators,
\begin{align}
AD_{0} &=
\begin{pmatrix}
1 & 0 \\
0 & \sqrt{1 - p_d - p_a}
\end{pmatrix} = \frac{1+\sqrt{1 - p_d - p_a}}{2}I+\frac{1-\sqrt{1 - p_d - p_a}}{2}Z, \notag\\
AD_{1} &=
\begin{pmatrix}
0 & 0 \\
0 & \sqrt{p_d}
\end{pmatrix}= \frac{\sqrt{p_d}}{2}I - \frac{\sqrt{p_d}}{2}Z, \notag\\
AD_{2} &=
\begin{pmatrix}
0 & \sqrt{p_a} \\
0 & 0
\end{pmatrix} = \frac{\sqrt{p_a}}{2}X+\frac{i\sqrt{p_a}}{2}Y, \label{ADkraus}
\end{align}
where $X$ and $Y$ are Pauli operators and
\begin{align}
p_a &= 1 - e^{-t/T_1}=\gamma, \notag\\
p_d &= e^{-t/T_1} - e^{-2t/T_2}=4\lambda(1-\lambda)-\gamma, \notag\\
\sqrt{1 - p_p - p_a} &= e^{-t/T_2}=1-2\lambda.
\end{align}
\subsection{Drift}
\par
Frequency drift, often referred to as detuning noise, arises from slow temporal fluctuations in the qubit transition frequency caused by variations in the local electromagnetic environment, control electronics, or residual couplings to other degrees of freedom. In superconducting qubits, such drift is commonly induced by flux noise, charge noise, or calibration imperfections, leading to coherent but unintended phase accumulation during idle periods. Unlike amplitude damping and dephasing, detuning noise is fundamentally a coherent error, although ensemble averaging over fluctuations can result in effective dephasing.
\par
The effect of detuning can be modeled by an additional term in the qubit Hamiltonian \cite{Krantz:2019jkw}
\begin{align}
    H_{\mathrm{drift}} = \frac{\delta\omega}{2} Z ,
\end{align}
where $\delta \omega$ denotes the frequency offset from the intended qubit resonance, commonly referred to as the detuning or drift frequency, or simply detuning. For static or quasi-static detuning, this Hamiltonian generates a unitary evolution during an idling period of duration $t$
\begin{align}
    U_{\mathrm{drift}}(t) = \exp\!\left(- i \frac{\delta\omega t}{2} Z \right),
    \label{drifkraus}
\end{align}
which corresponds to an unintended rotation about the Z-axis of the Bloch sphere. For a fixed detuning frequency $\delta \omega$ the evolution of a density matrix $\rho$ is given by
\begin{align}
    \mathcal{E}_{\mathrm{drift}}(\rho)= U_{\mathrm{drift}}\, \rho \, U_{\mathrm{drift}}^\dagger =
    \begin{pmatrix}
\rho_{00} &
e^{- i \delta\omega t}\rho_{01} \\
e^{ i \delta\omega t}\rho_{10} &
\rho_{11}
\end{pmatrix}.
\end{align}
This expression shows that frequency drift leaves population terms unchanged while inducing a deterministic phase rotation in the off-diagonal coherence terms.
\par
In practice, the detuning $\delta \omega$ is often not constant but varies slowly over time or across experimental repetitions. If $\delta \omega$ is modeled as a random variable with zero mean and variance $\sigma^2$, averaging over realizations leads to an effective decay of coherence,
\begin{align}
    \langle \rho_{01}(t) \rangle
= \rho_{01}(0)\, \langle e^{- i \delta\omega t} \rangle .
\end{align}
which, for Gaussian-distributed detuning, yields
\begin{align}
    \langle \rho_{01}(t) \rangle
= \rho_{01}(0)\, e^{-\frac{1}{2}\sigma^2 t^2}.
\end{align}
This Gaussian decay can be viewed as the low-frequency end of the dephasing noise spectrum and contrasts with the exponential decay associated with Markovian dephasing.
\par
Frequency drift thus represents an important source of idling error that bridges coherent control errors and incoherent noise. Its impact depends strongly on the duration of idle periods and the scheduling of gate operations, making it particularly relevant in the context of circuit-level timing optimization and error suppression strategies.
\\
\par
We note that additional noise sources such as residual ZZ interactions, crosstalk, and leakage, are not explicitly incorporated into the present model, and their impact on the effectiveness of timing-based error suppression remains to be systematically investigated.
\section{Superconducting device and CNOT gate}\label{2methods}
\par
\begin{figure}[bht]  
    \centering  
    \includegraphics[width=50mm]{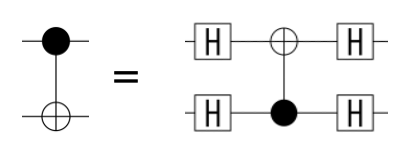} 
    \caption{\centering A reversed CNOT gate.}  
    \label{CNOT}  
\end{figure}

The controlled-NOT (CNOT) gate, or more generally the cross-resonance (CR) gate, is part of the standard gate set of quantum processors and plays a central role in generating entanglement between qubits. In superconducting quantum devices, a widely adopted implementation of the CNOT gate exploits the cross-resonance interaction between fixed-frequency qubits \cite{Kwon:2021oxt,Krantz:2019jkw,Chow:2011zkc,Rigetti2010}. In this case, using the higher-frequency qubit as the control and the lower-frequency qubit as the target significantly reduces the gate execution time compared to the reverse configuration. As a result, superconducting platforms typically impose a hardware-level constraint that the higher-frequency qubit must act as the control and the lower-frequency qubit as the target. However, this constraint is not explicitly enforced at the software level, allowing circuit designers to specify CNOT gates with the opposite control direction. To reconcile this discrepancy between logical circuit descriptions and physical hardware constraints, circuit compilers and microwave pulse compilers transform such disallowed CNOT gates into equivalent operations that are compatible with the device. Specifically, the control direction of a CNOT gate can be reversed by inserting H gates before and after the operation. As illustrated in Fig. \ref{CNOT}, applying H gates to both qubits enables the logical realization of a CNOT gate in the opposite direction using a physically permitted CNOT implementation.
\par
The H gates contained in such CNOT constructions often possess scheduling flexibility in quantum circuits. In other words, when idle periods exist for certain qubits before or after a reversed CNOT gate, the H gates need not be applied immediately before and after the CNOT gate; in the absence of noise, applying them at any point within the idle interval yields an equivalent operation. However, because the probability of idling errors depends on time, the placement of the H gates influences the accumulation of idling errors and consequently affects the accuracy of the computational results.
\par
Motivated by this observation, this work proposes two methods for suppressing idling errors by appropriately adjusting the execution timing of the H gates associated with CNOT gates. Unlike dynamical decoupling, these approaches do not require the insertion of additional gate operations, rendering them effective even on quantum devices with limited gate fidelity.

\subsection{Method 1: Midpoint placement}
Rule: for every H gate that must occur within an idling interval, place it at the midpoint of that interval.
\par
Consider an unknown single-qubit state. Depending on whether the state is closer to the computational basis states $\{|0\rangle,|1\rangle\}$ or the superposition basis states $\{|+\rangle,|-\rangle\}$, it becomes effectively more sensitive to either amplitude damping or dephasing, respectively. Placing the H gate entirely at the beginning or end of the idle interval can therefore produce a “worst-case” configuration in which one of these error mechanisms acts over the entire idle duration in the most vulnerable basis. In contrast, placing the H gate at the midpoint of the idle interval distributes the exposure more evenly between the two basis orientations induced by the H gate, thereby mitigating the dominance of either error mechanism.

\subsection{Method 2: Random placement}
Rule: instead of fixing the position of the H gate, sample it randomly under an uniform distribution over the idle interval for each shot, execute the circuit accordingly, and average the measurement outcomes over all shots.
\par
let $\mathcal{E}_t$ denote the operation when the operation when the H gate is placed at time $t\in [0,T]$. The randomized strategy implements the mixture channel
\begin{align}
\mathcal{E}
&=\frac{1}{T}\int_{0}^{T}\mathcal{E}_t\,dt.\label{randchannel}
\end{align}
Quasi-static detuning and slow phase noise often manifest as a small unwanted $Z$-rotation whose angle depends on idle time:
\begin{align}
U_{\theta(t)} &= e^{i\theta(t) Z}, \qquad |\theta(t)|\ll 1.
\end{align}
To first order,
\begin{align}
U_{\theta(t)}\,\rho\,U_{\theta(t)}^\dagger
&=
\rho + i\theta(t)\,[Z,\rho] + O(\theta(t)^2).
\end{align}
If the timing dependence $\theta(t)$ has an approximate symmetry about the midpoint — e.g., the induced angle is antisymmetric around $T/2$ - then averaging over $t$ suppresses (at least a part of) the first-order coherent contribution:
\begin{align}
\frac{1}{T}\int_{0}^{T}\theta(t)\,dt \approx 0
\quad\Rightarrow\quad
\mathcal{E}(\rho) \approx \rho + O(\theta^2).
\end{align}
In that case, random placement converts a coherent timing-dependent phase error into a weaker and effectively higher-order residual.
\section{Time evolution and trace distance}
\par
\begin{figure}[bt]  
    \centering  
    \includegraphics[width=50mm]{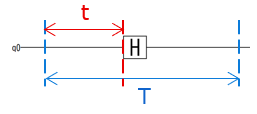} 
    \caption{\centering a single-qubit circuit with a total idling time $T$, during which a single H gate is applied at an intermediate time $t$}  
    \label{idlec}  
\end{figure}

To further investigate the effects of different scheduling timings, we introduce the time evolution of the density matrix and the corresponding trace distance. Here we consider a single-qubit circuit with a total idling duration $T$, during which a single H gate is applied at an intermediate time $t$ (see Fig.~\ref{idlec}). The qubit undergoes idling error both before and after the gate operation. Our goal is to determine the optimal gate timing that minimizes the overall effect caused by idling error.
\par
Let the initial state be described by the density matrix $\rho(0)$. Using (\ref{ADkraus}) and (\ref{drifkraus}), the final output state at $T$ is given by
\begin{align}
\rho(T,t)
=
\sum_{i,j=0}^{2}
&AD_{_j}(T-t)
\, e^{-i\omega (T-t) Z/2}
\, H \ AD_{_i}(t)\, e^{-i\omega t Z/2}
\, \rho(0)
\, e^{i\omega t Z/2}
\, AD_{_i}^{\dagger}(t)\notag\\
&H
\, e^{i\omega (T-t) Z/2}
\, AD_{_j}^{\dagger}(T-t).
\label{finalstate}
\end{align}
In the absence of idling noise, the ideal final state would be
\begin{equation}
\rho' = H \rho(0) H.\label{idealstate}
\end{equation}
To quantify the deviation from the ideal state, we use the trace distance
\begin{align}
D(\rho, \sigma)
&=
\frac{1}{2} \|\rho - \sigma\|_1
=
\frac{1}{2} \operatorname{Tr} |\rho - \sigma|,
\end{align}
where the trace norm is defined as
\begin{align}
\|A\|_1 = \operatorname{Tr}|A|, \quad |A| = \sqrt{A^\dagger A}.
\end{align}
Unlike fidelity, which is defined through the overlap between density matrices, the trace distance is computed from the difference of each component of the matrices. As a consequence, it is more sensitive to differences in both populations (diagonal components) and coherences (off-diagonal components). In this work, we evaluate the trace distance $D\bigl(\rho(T,t), \rho'\bigr)$ as a function of the gate timing $t$. The optimal timing $t^\ast$ is defined as
\begin{equation}
t^\ast
=
\arg\min_{t \in [0,T]}
D\bigl(\rho(T,t), \rho'\bigr).
\end{equation}
By appropriately selecting $t^\ast$, we determine the gate operation timing that minimizes the impact of idling errors over the total duration $T$. This optimization provides a practical approach for suppressing decoherence through temporal control alone, without modifying the underlying physical noise parameters of the system or introducing additional gate operations.
\section{Results}
In this section, we validate the two timing heuristics introduced in Sec.~\ref{2methods} using (i) hardware experiments on the 64 qubit superconducting quantum computer at the RIKEN RQC-Fujitsu Collaboration Center for midpoint placement with fixed initial states and (ii) numerical simulations by QuTiP\cite{qutip5,Li2022pulselevelnoisy} for both midpoint placement and random placement, with fixed as well as Haar random initial states.
\par
The purpose of the experimental study is not to demonstrate improvements at large-scale circuit level, but rather to isolate and verify the fundamental interplay between idling noise processes and gate scheduling predicted by our theoretical analysis.
\subsection{Experiment}
\begin{figure}[bt]  
    \centering  
    \includegraphics[width=130mm]{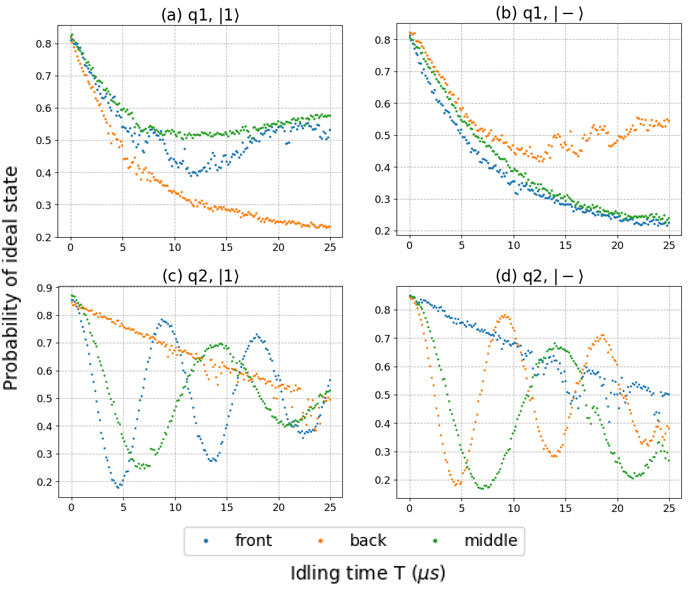} 
  \caption{Single-qubit idling experiment results. (a) q1 with initial state $|1\rangle$. (b) q1 with initial state $|-\rangle$. (c) q2 with initial state $|1\rangle$. (d) q2 with initial state $|-\rangle$.}
  \label{fig:1qex}
\end{figure}

We first implemented the single‑qubit idling circuit shown in Fig.~\ref{idlec} . For a fixed idling interval of duration $T$, we executed three circuits that differ only in the placement timing of the H gate within the idle window:
\begin{itemize}
\item front: H gate placed at the beginning of the idling interval,
\item middle: H placed at the center of the idling interval,
\item back: H placed at the end of the idling interval.
\end{itemize}
At the end of each circuit, we measured the output state probabilities. As initial states, we used both $|1\rangle$ and $|-\rangle$, as the behaviors for $|0\rangle$ and $|+\rangle$ are comparatively straightforward. The experiments were performed on two representative physical qubits (labeled as q1 and q2) of the device, with the following measured coherence parameters:\\
q1: $\ T_1=5.24~\mu s,\ T_2=8.44~\mu s,\ \omega=0.06~$MHz\\
q2: $\ T_1=32.81~\mu s,\ T_2=43.34~\mu s,\ \omega=0.11~$MHz\\
Here, $\omega$ denotes the effective detuning frequencies. Each data point was estimated from 10,000 shots. The results are shown in Fig.~\ref{fig:1qex}. The y-axis denotes the probability of obtaining the ideal state in the absence of idling errors (e.g., $|-\rangle$ for initial state $|1\rangle$) and the x-axis denotes the total idling duration. We see that for idling durations sufficiently short such that drift-induced oscillations are not yet dominant (e.g., over the entire range shown in (a) and (b), and for durations $\lessapprox 6\mu s$ in (c)and (d)), midpoint placement consistently achieves the best or near-best performance. This result supports the key claim of Method 1: placing the H gate at the midpoint provides a robust default strategy that avoids worst-case performance in the practically relevant idling regime.



\par
\begin{figure}[bt]  
    \centering  
    \includegraphics[width=130mm]{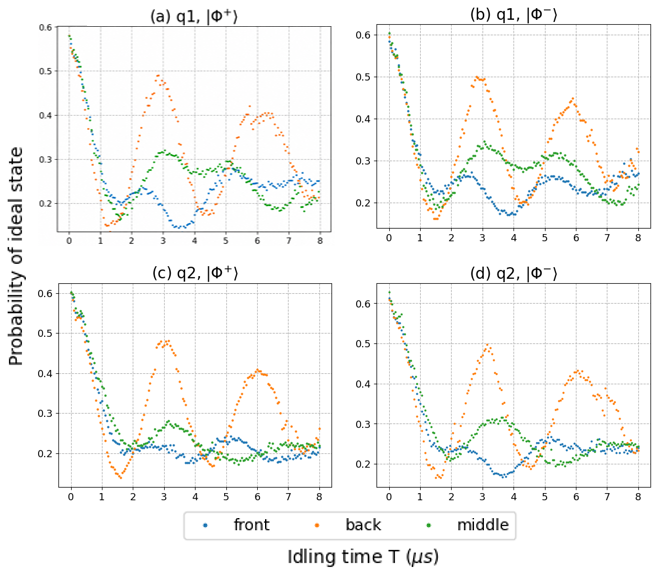} 
  \caption{Two-qubit idling experiment results. The y-axis denotes the probability of obtaining the ideal state with no idling error occurs. (a) H gate acts on q1 with initial state $|\Phi^{+}\rangle$. (b) H gate acts on q1 with initial state $|\Phi^{-}\rangle$.  (c) H gate acts on q2 with initial state $|\Phi^{+}\rangle$.  (d) H gate acts on q2 with initial state $|\Phi^{-}\rangle$. }
  \label{fig:2qex}
\end{figure}

Next, we extend the analysis beyond the single-qubit and perform an analogous idling experiment using two-qubit entangled states. We first prepare the Bell states as initial states and then introduces an idling window during which a H gate is applied to one of the qubits using front, middle, or back placement timing. Here we used two types of Bell states defined as
\begin{align}
|\Phi^{\pm}\rangle &= \frac{|00\rangle \pm |11\rangle}{\sqrt{2}} .
\end{align}
From Fig.~\ref{fig:2qex}, the two-qubit results show the same qualitative behavior as the single-qubit experiments. In particular, for sufficiently short idling intervals (e.g., $\lessapprox 1.5\mu s$ in these experiments), midpoint placement again avoids the worst-case behavior, while back placement yields the poorest performance. These observations support the idea that midpoint placement remains a robust heuristic even for multi-qubit state inputs which contain entanglements.


\par
Although the experiments were conducted on single- and two-qubit circuits, the underlying mechanism—namely, the redistribution of idle-time exposure under noise—is fundamentally independent of system size. Consequently, the proposed scheduling strategies are expected to generalize to larger quantum circuits as well.
\subsection{Simulation}
\subsubsection{Fixed initial states}
To gain further insight into the mechanisms underlying the observed behaviors, we performed numerical simulations using QuTiP. Two physical models were considered:
a Linear Spin Chain (LSC) model , and an SCQubits-based (SCQubits) model tailored to superconducting devices. In both models, we included dissipative noise consistent with measured $T_1$ and $T_2$, and an additional drift term proportional to Pauli $\sigma_z$ in the Hamiltonian, representing an effective frequency detuning.
\par
\begin{figure}[bt]  
    \centering  
    \includegraphics[width=130mm]{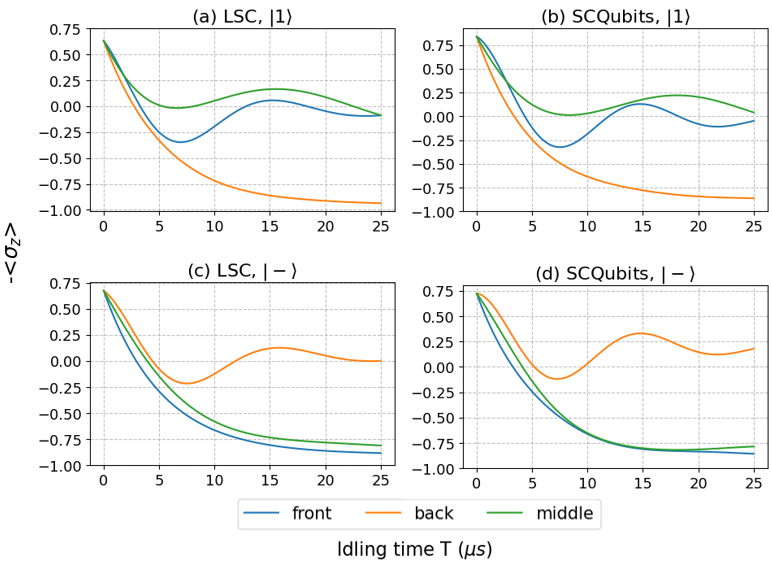} 
  \caption{Single-qubit idling simulation results for q1. (a) LSC model with initial state $|1\rangle$. (b) SCQubits model with initial state $|1\rangle$. (c) LSC model with initial state $|-\rangle$. (d) SCQubits model with initial state $|-\rangle$.}
  \label{fig:1qsimq1}
\end{figure}
\begin{figure}[bt]  
    \centering  
    \includegraphics[width=130mm]{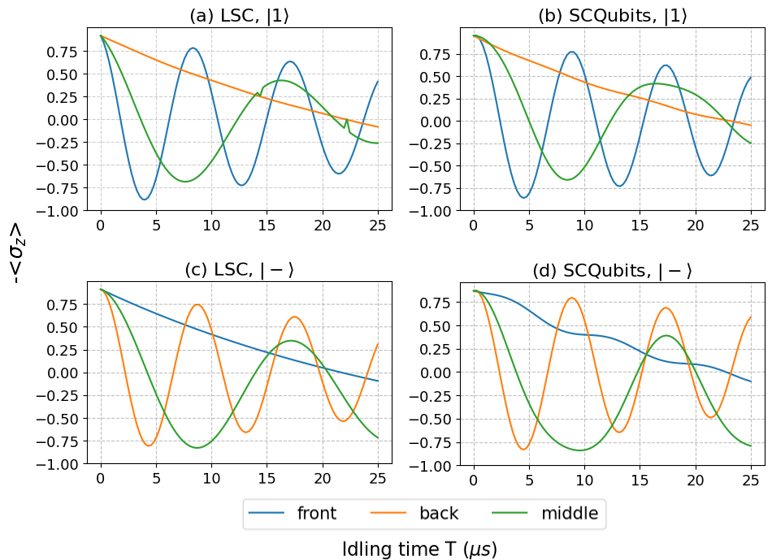} 
  \caption{Single-qubit idling simulation results for q2. (a) LSC model with initial state $|1\rangle$. (b) LSC model with initial state $|-\rangle$. (c) SCQubits model with initial state $|1\rangle$. (d) SCQubits model with initial state $|-\rangle$.}
  \label{fig:1qsimq2}
\end{figure}

The single-qubit simulation results are shown in Fig.~\ref{fig:1qsimq1} and Fig.~\ref{fig:1qsimq2}. Here we computed $\langle \sigma_z \rangle$ (or $-\langle \sigma_z \rangle$ for easier visual comparison with the experimental results) of the output states, which is directly related to the measured computational-basis population. Both models successfully reproduce the overall experimental trends across front/middle/back placements. However, the SCQubits model better captures the oscillatory features observed in the hardware data, including the notably small oscillations in the front/back placements (e.g. in Fig.~\ref{fig:1qex}(d) and Fig.~\ref{fig:1qsimq2}(d) for the front placement), suggesting it provides a more accurate description of the device behavior.


%


\par
\begin{figure}[bt]  
    \centering  
    \includegraphics[width=130mm]{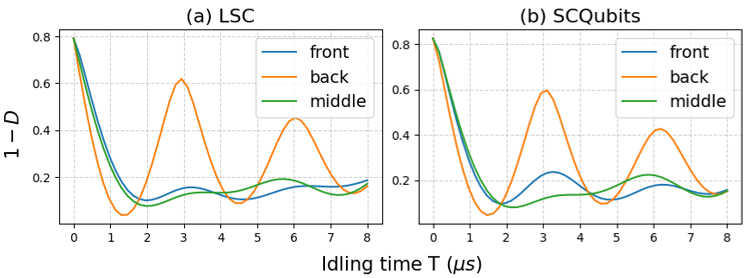} 
  \caption{Two-qubit idling simulation results with H gate acting on q2 and initial state $|\Phi^{+}\rangle$. (a) LSC model. (b) SCQubits model.}
  \label{fig:2qsim}
\end{figure}

For the two-qubit case, we evaluated simulation accuracy using the trace distance between the input and output density matrices, and the results are plotted in Fig.~\ref{fig:2qsim}. To facilitate visual comparison across different plots, we use $1-D$ as the y-axis, where $D$ denotes the trace distance. Both models reproduce the experimental behavior to some extent; however, an important discrepancy emerges. In the LSC model, the midpoint placement never becomes the optimal pattern for any tuning of the drift coefficient, in contrast to the experimental results (see $\lessapprox 2\mu s$ in Fig.~\ref{fig:2qsim} and (c), (d) in Fig.~\ref{fig:2qex}). This indicates that accurately capturing multi-qubit idling behavior requires a model more closely aligned with the physics of superconducting devices, thereby motivating the use of the SCQubits model for the analysis of multi-qubit dynamics. A further observation is that, in both models, reproducing the two-qubit behavior requires larger effective drift coefficients than those used in the single-qubit simulations. This discrepancy is likely attributable to the enhanced influence of drift arising from two-qubit operations.

\par
We next evaluate the random placement method using the SCQubits model. As in the midpoint placement, we consider both single-qubit and two-qubit idling circuits, but instead of fixing the timing to the front/middle/back, we randomize (or equivalently, uniformly average) the insertion timing of the H gate within a fixed idling window.
\par
For a given idling duration $T$, we discretize the window into $N=100$ equally spaced time points $\{t_k\}^N_{k=1}\subset [0,T]$. For each $t_k$, we run the same circuit with the H gate inserted at $t=t_k$, and then compute the time-averaged output state \begin{align}
\rho_{\mathrm{rand}}(T)
&\approx \frac{1}{N}\sum_{k=1}^{N}\rho(T,t_k),
\end{align}
which approximates the idealized random-placement channel (\ref{randchannel}).
\par
\begin{figure}[bt]
  \centering
  \subfloat[]{
    \includegraphics[width=0.49\columnwidth]{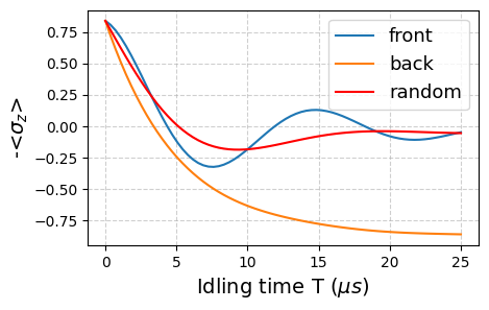}
  }
  \subfloat[]{
    \includegraphics[width=0.465\columnwidth]{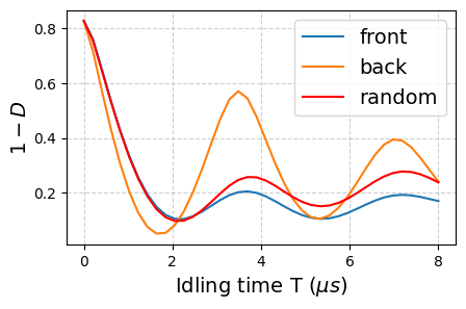}
  }
  \caption{Simulation results with random placement. (a) Single-qubit circuit with initial state $|1\rangle$. (b) Two-qubit circuit with initial state $|\Phi^{+}\rangle$.}
  \label{fig:randsim}
\end{figure}

We apply this procedure to both the single-qubit and two-qubit initial states used in midpoint placement. As representative examples, we show the results for the initial states $|1\rangle$ and $|\Phi^{+}\rangle$ in Fig.~\ref{fig:randsim}. Overall, random placement exhibits similar qualitative behavior as midpoint placement: within the practically relevant range of idling durations, random placement does not become the worst-performing strategy and generally matches or outperforms the edge placements (see Fig.~\ref{fig:randsim} in comparison with Fig.~\ref{fig:1qsimq1} (b) and Fig.~\ref{fig:2qsim} (b)).
\subsubsection{Random initial states}
In practical quantum computation, it is generally difficult to know the exact quantum state of a qubit at every point within a circuit. Consequently, it is not possible to select the optimal placement strategy conditioned on the actual input state during computation. To evaluate the proposed timing strategies under such uncertainty, we consider the scenario in which the pre-H state is unknown and randomly distributed.
\par
To this end, we generate multiple Haar-random pure states (see Appendix\ref{Haarrand} for details) as initial states and evaluate the idling circuit for each sample. For every realization, we compute the trace distance between the ideal and noisy output density matrices and subsequently average the results over all samples. It should be emphasized that the Haar-random averaging should not be viewed as a realistic model of the state distribution in practical quantum computations, but rather as a state-agnostic benchmark that eliminates bias toward any particular basis or class of states. While the Haar-averaged results provide a useful state-independent criterion, evaluating the proposed strategies under structured state ensembles relevant to specific quantum algorithms remains an important direction for future work.
\par
\begin{figure}[bt]
  \centering
  \subfloat[]{
    \includegraphics[width=0.48\columnwidth]{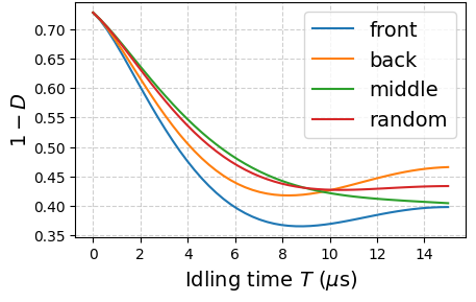}
  }
  \subfloat[]{
    \includegraphics[width=0.48\columnwidth]{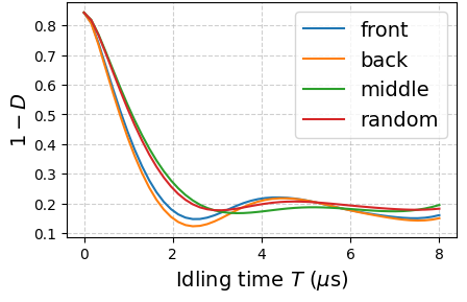}
  }

  \caption{Simulation results with Haar random initial states. The y-axis is chosen to be $1-D$ also for the single-qubit case. (a) Single-qubit (q1) circuit. (b) Two-qubit circuit with H gate acting on q2.}
  \label{fig:haarrandsim}
\end{figure}

Let $\rho_s(0)$ be the s-th sampled initial state and $\rho_s(T)$ the corresponding output state under a given placement rule. The averaged metric is
\begin{align}
\overline{D}(T)
&= \frac{1}{N}\sum_{s=1}^{N} D\!\left(\rho_s(T),\rho_s(0)\right).
\end{align}
And the results are plotted in Fig.~\ref{fig:haarrandsim}.
\par
The number of samples is set to 100 for single-qubit states and 50 for two-qubit states. For both the single-qubit and two-qubit cases, when the pre-H state is unknown, midpoint placement and random placement yield the smallest average trace distance over a reasonable range of idling durations, suggesting that the two strategies minimize the average impact of idling noise. Moreover, the difference between midpoint placement and random placement is generally small in the averaged results. As an example, for the single-qubit case, an improvement of up to 12\% in accuracy compared to the unfavorable placements is observed at $T=5 \mu s$. The range of idling durations over which midpoint and random placement remain advantageous depends on device-specific parameters such as the qubit coherence times $T_1,\ T_2$ and the drift strength. Given that typical idling durations in quantum circuits are on the order of several microseconds, the proposed timing strategies are expected to be effective across a broad range of practical quantum computations.
\par
Finally, beyond the averaged results, we also observe robustness at the per-sample level: even without averaging over random states, neither midpoint placement nor random placement was observed to yield the worst-case performance among the tested strategies across the sampled states, within the experimentally relevant idling-time range.

\section{Analytic computation}
In this section, we provide a further theoretical analysis of the qubit time evolution and and detailed explanations for the experimental and simulation results discussed above.
\subsection{Drift only}
To obtain analytical insight into the timing dependence, we first consider the simplified case in which only coherent drift (detuning) is present, and no amplitude damping or dephasing occurs. The qubit experiences a constant (quasi-static) detuning $\omega$ over the full time window of interest. In this setting, the idling evolution is unitary and corresponds to a Z-axis rotation whose angle grows linearly with elapsed time
\begin{align}
    \rho(T,t)
=
e^{-i\frac{\omega}{2}(T-t)Z}
\, H \,
e^{-i\frac{\omega}{2}tZ}
\, \rho(0) \,
e^{i\frac{\omega}{2}tZ}
\, H \,
e^{i\frac{\omega}{2}(T-t)Z}.\label{rhodrift}
\end{align}
We compute the trace distance between the noisy final state and the ideal state, and then average over Haar-random single-qubit inputs. Working in the Pauli basis, we obtain a closed-form expression for the Haar integral of $\overline{D}^2$\footnote{Here we compute $\overline{D}^2$ instead of $\overline{D}$ for computational convenience. Given that $\overline{D}\in [0,1]$, the minimal timings of the Haar averaged $\overline{D}$ and $\overline{D}^2$ are considered the same}, up to an overall proportionality constant set by normalization of the Haar measure
\begin{align}
    \int d\mu_2 \, D^2
\;\propto\;
8 - 2\bigl(\cos(\omega t)+1\bigr)\bigl(\cos(\omega (T-t))+1\bigr). \label{Ddrift2}
\end{align}
Expanding this expression yields
\begin{align}
    \int d\mu_2 \, D^2
\;\propto\;
6-\Bigl(\cos(\omega T)+\cos(\omega(T-2t))
+4\cos\!\bigl(\tfrac{\omega T}{2}\bigr)\cos\!\bigl(\tfrac{\omega (T-2t)}{2}\bigr)\Bigr).\label{Ddrift}
\end{align}
The resulting expression is a composite wave of multiple frequency components in $t$. It typically reaches larger values near the endpoints $t=0$ and $t=T$, and becomes smaller near the center of the interval, depending on the accumulated phase $\omega T$. Note that the trace distance decreases as the bracketed term in (\ref{Ddrift}) increases, and all the bracketed terms are positive in the range $\omega T \leq \pi/2$. Therefore it is natural to consider the existence of a minimal point, which likely locates at the center $t=T/2$.
\par
\begin{figure}[bt]  
    \centering  
    \includegraphics[width=110mm]{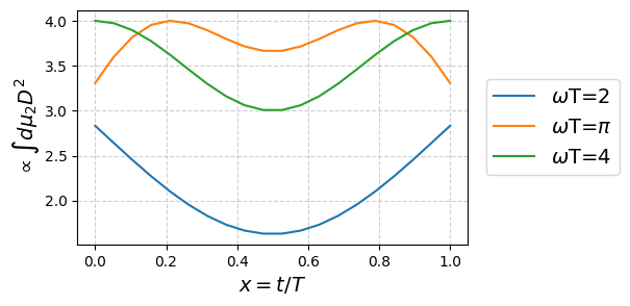} 
    \caption{\centering Eq.~(\ref{Ddrift2}) with various values of $\omega T$.}   
    \label{fig:wTvary} 
\end{figure}

To study its structure more transparently, we introduce the normalized variable
\begin{align}
x \equiv t/T  \in [0,1].\label{xdef}
\end{align}
Substituting into (\ref{Ddrift2}), we find that for small accumulated phase $\omega T \leq \pi/2$, this function is convex and attains its minimum at $x=1/2$, which suggests that the trace distance is minimized at the center of the idling interval. However, when the accumulated phase becomes sufficiently large, the function develops additional oscillatory structure and the midpoint no longer corresponds to the global minimum. Thus, the midpoint scheduling is optimal only in the small-phase regime. We verified this analytical behavior by plotting (\ref{Ddrift2}) with various values of $\omega T$ (Fig.~\ref{fig:wTvary}).
\par
This “midpoint” rule can be interpreted as a simple symmetry-based mitigation of coherent phase accumulation: the detuning-induced phase errors accumulated before and after the gate interfere in a way that reduces the average disturbance when the idling interval is temporally balanced.

\subsection{Amplitude damping and dephasing}
\begin{figure}[bt]  
    \centering  
    \includegraphics[width=140mm]{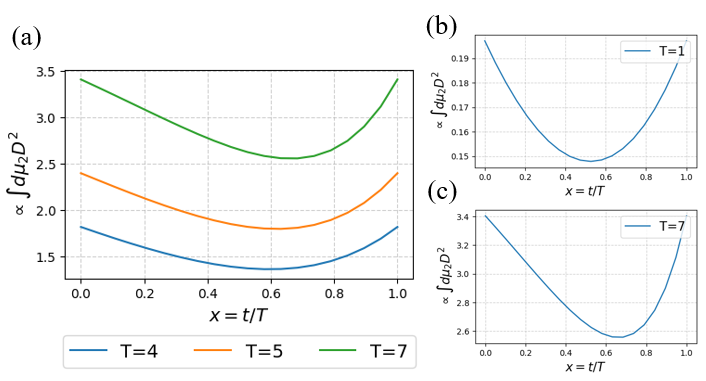} 
    \caption{\centering (a) Eq.~(\ref{DT1T2}) with various values of $T \ (\mu s)$. (b) $T=1(\mu s)$. (c) $T=7(\mu s)$}   
    \label{fig:Tvary} 
\end{figure}

We now extend the analysis to the case where both amplitude damping and dephasing are present. The noise channel is described by the combined Kraus operators introduced earlier in (\ref{ADkraus}). The final density matrix is then given by
\begin{align}
\rho(T,t)
=
\sum_{i,j=0}^{2}\,
AD_j(T-t)\,
H\,
AD_i(t)\,
\rho(0)
AD_i^\dagger(t)\,
H\,
AD_j^\dagger(T-t).\label{rhoAD}
\end{align}
Again integrating the trace distance over the Haar measure yields
\begin{align}
\int d\mu_2\, D^2
\propto
(a-aB-1)^2
+
(A-Ab-1)^2
+
(1-aA)^2
+
3A^2 b^2
+
3B^2,\label{DT1T2}
\end{align}
where
\begin{align}
a = e^{-t/T_2},\quad
A = e^{-(T-t)/T_2},\quad
b = 1-e^{-t/T_1},\quad
B = 1-e^{-(T-t)/T_1}.
\end{align}
Note that all terms except for the last two are symmetric about $t=T/2$. And the term $(1-aA)^2$ depends only on the total duration $T$ and is independent of $t$. The last two terms reflect the non-commutativity between relaxation and the Hadamard operation. Because H gate exchanges Z- and X-axes, relaxation occurring before and after the gate affects different Bloch components, producing a generally asymmetric dependence on $t$. Due to partial symmetry in the expression, $t=T/2$ is always a stationary point of the symmetric contributions. However, the asymmetric terms can shift the global minimum away from the midpoint when decoherence is strong. To further analyze (\ref{DT1T2}), we again 
use the normalized variable (\ref{xdef}). Differentiating (\ref{DT1T2}) with respect to $x$ and expanding around small $T$, we obtain
\begin{align}
\Big( \int d\mu_2\, D^2 \Big)'
\propto\,&
\ \frac{T(2x-1)\bigl(T_1^2-2T_1T_2+4T_2^2\bigr)}{T_1^2T_2^2}-\frac{3T^2}{2T_1^2T_2^3}\Bigl[
(2x-1)(T_1^3+4T_2^3)
\notag\\
&+T_1^2T_2(1-2x)+T_1T_2^2(-6x^2+2x+1)
\Bigr]
+O(T^3).
\end{align}
The leading order term in $T$ is proportional to $2x-1$, hence it implies a unique minimum at $x=1/2$. However, higher-order contributions introduce additional $x$-dependence, which breaks the exact midpoint optimality beyond leading order. By plotting (\ref{DT1T2}) for various values of $T$ (Fig.~\ref{fig:Tvary}), we see that for sufficiently small $T$, (\ref{DT1T2}) is nearly symmetric and minimized close to $x=1/2$, while for larger $T$ the minimum shifts slightly to the right side from the midpoint.
\par
Thus, unlike the purely coherent drift case, dissipative relaxation introduces a competition between pre- and post-gate decay, and the optimal timing depends quantitatively on the relative magnitudes of $T_1$, $T_2$ and $T$.

\subsection{$T_1 T_2$ decoherence and drift}
Finally, we consider an idling model that incorporates both amplitude damping/dephasing and quasi-static detuning. By computing the output state in (\ref{finalstate}) and subsequently evaluating the Haar-averaged trace distance, we obtain
\begin{align}
\int d\mu_2\, D^2 \;\propto
&-2A(1-b)\cos\bigl(\omega(T-t)\bigr)
-2a(1-B)\cos(\omega t)-2Aa\cos(\omega t)\cos\bigl(\omega(T-t)\bigr)\notag\\
&+A^2a^2
+3+A^2(1-b)^2
+a^2(1-B)^2+3A^2b^2+3B^2
.\label{Dallnoise}
\end{align}
This Haar-averaged trace distance decomposes naturally into a midpoint-symmetric part and a generally asymmetric part: all term are symmetric under $t\mapsto T-t$ except for the last two terms (which are identical to those discussed in the previous section). Consequently, $t=T/2$ remains a robust “first guess” for minimizing the Haar-averaged trace distance. 
\par
\begin{figure}[bt]
  \centering
    \includegraphics[width=0.55\columnwidth]{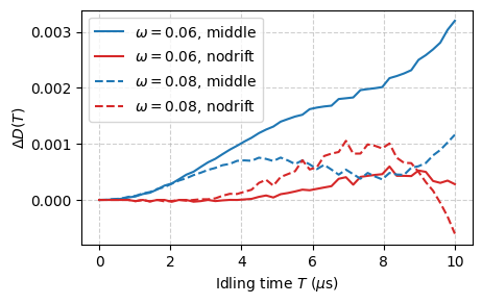}
  \caption{Differences of trace distance $\Delta D(T)= D_{\mathrm{middle/nodrift}}(T) - D_{\mathrm{min}}(T)$. The values always remain positive in the practically relevant idle regime.}
  \label{fig:idealsim}
\end{figure}

To further quantify the performance difference between the exact minimum placement and the midpoint placement, we compute the differences of the corresponding Haar-averaged trace distances with several drift strengths $\omega$ through QuTiP simulations and plot the results in Fig.~\ref{fig:idealsim}. The vertical axis is defined as $\Delta D(T)= D_{\mathrm{middle/nodrift}}(T) - D_{\mathrm{min}}(T)$, where $D_{\mathrm{middle}}(T)$, $D_{\mathrm{min}}(T)$ and $D_{\mathrm{nodrift}}(T)$ denote the trace distance evaluated with the midpoint placement, the exact minimum timing placement and the minimum timing placement but with the drift strength set to $0$ in (\ref{Dallnoise}), respectively. The number of Haar-random state samples is set to $6000$. Fig.~\ref{fig:idealsim} shows that in the regime where the drift-induced oscillation is not yet dominant, $\Delta D(T)$ remains positive, indicating that the drift-aware minimum placement yields smaller trace distance. Additionally, smaller drift makes the minimal timing deviate more from the midpoint. This is consistent with the analytic structure: the drift-dependent terms are symmetric about $t=T/2$, therefore when drift is weak, non-drift contributions play a larger role in shifting the minimizer away from the midpoint. 
\par
Nevertheless, the differences remain below $1\%$ in all cases within a practically reasonable idling length. These results therefore clarify the role of the midpoint rule: midpoint placement is a robust, calibration-free default that performs near-optimally in the experimentally relevant idle regime, while exact minimum placement can provide additional improvement, particularly for longer idling intervals or when a reliable device parameter (including effective drift) estimate is available.

\subsection{Random placement}
\subsubsection{An pedagogical example}
Before proceeding to Haar-random inputs, it is instructive to analyze a minimal setting that makes the mechanism of timing randomization transparent. Here we consider the case of an initial state $|1\rangle$ under a drift-only noise model. Since $|1\rangle$ is an eigenstate of $Z$, the only meaningful effect of drift occurs after the H gate, during the remaining interval $(T-t)$. In this special case, the output state is a pure state obtained by applying a Z-rotation with angle $\omega(T-t)$ to the ideal target state $|-\rangle =H |1\rangle$. The corresponding density matrix can be written as
\begin{align}
\rho(T,t)
&=
\frac{1}{2}
\begin{pmatrix}
1 & -e^{i\omega(T-t)}\\
-e^{-i\omega(T-t)} & 1
\end{pmatrix},\label{rhoori}
\end{align}
the trace distance to the ideal output $\rho_{\mathrm{id}}=|-\rangle\!\langle -|$ is given by
\begin{align}
D(t)
&=
D\bigl(\rho(T,t),\rho_{\mathrm{id}}\bigr)
=
\left(\frac{1}{2}\left[1-\cos\!\bigl(\omega(T-t)\bigr)\right]\right)^{1/2}.
\end{align}
For small accumulated phase $\omega T \ll 1$, this behaves as
\begin{align}
D(t) \approx \frac{|\omega|}{2}(T-t).
\end{align}
Hence, for front placement $t=0:\ D \approx \frac{\omega T}{2}$ and for middle placement $t=\frac{T}{2}:\ D \approx \frac{\omega T}{4}$. This shows that placing H gate at the middle can reduce the leading coherent contribution by a factor of two relative to the front.
\par
To illustrate the random-placement method, we now sample $t$ uniformly in $[0,T]$ for each shot and average the resulting outcomes. At the state level, this corresponds to averaging the density matrix:
\begin{align}
\rho_{\mathrm{rand}}(T) &= \frac{1}{T}\int_0^T \rho(T,t)\,dt .\label{rhorand}
\end{align}
Next, we focus on the off-diagonal components of the density matrix, since only the off-diagonal phase factor depends on $t$. Defining
\begin{align}
\alpha
\equiv
\frac{1}{T}\int_0^T e^{i\omega(T-t)}\,dt ,
\end{align}
we can symmetrize the integral by the change of variables $t'=t-T/2$, yielding
\begin{align}
    \alpha&=
\frac{1}{T}e^{i\omega T/2}\int_{-T/2}^{T/2}
\Bigl(\cos(\omega t')-i\sin(\omega t')\Bigr)\,dt'\notag\\
&=\frac{2}{\omega T}e^{i\omega T/2}\sin\!\left(\frac{\omega T}{2}\right).
\end{align}
Here the second term in the first line is odd and cancels under symmetric integration. Therefore, (\ref{rhorand}) becomes
\begin{align}
\rho_{\mathrm{rand}}
&=
\frac{1}{2}
\begin{pmatrix}
1 & -\alpha\\[6pt]
-\alpha^* & 1
\end{pmatrix}.
\end{align}
The coherent factor $e^{i\omega T}$ in (\ref{rhoori}) is replaced by its average, resulting in a sin-like suppression in magnitude. Importantly, the integration removes the odd imaginary contribution of the integrand, which is precisely the mechanism by which random placement suppresses part of the first-order coherent contribution.
\par
Computing the trace distance to the ideal output gives
\begin{align}
D_{\mathrm{rand}}
&=
\left(
\frac{1}{4}
+\left(\frac{1}{\omega T}\right)^2\sin^2\!\left(\frac{\omega T}{2}\right)
-\frac{1}{2\omega T}\sin(\omega T)
\right)^{1/2}.
\end{align}
Expanding around $\omega T\to 0$,
\begin{align}
    D_{\mathrm{rand}}
=
\frac{\omega T}{4}
+O\bigl((\omega T)^3\bigr).
\end{align}
Thus it achieves the same leading-order scaling as midpoint placement: both reduce the small $\omega T$ trace-distance growth from $\omega T/2$ in the front placement to $\omega T/4$.
\subsubsection{Haar-random inputs}
The preceding example illustrates the central idea underlying random placement: by sampling the insertion time $t$ uniformly, timing-dependent coherent phases are partially canceled by averaging, converting a coherent, timing-sensitive error into a weaker effective disturbance. We now formalize this intuition at the state-agnostic level by computing the Haar-averaged error for (i) drift-only noise and (ii) 
$T_1 T_2$-only relaxation, and comparing random placement against other deterministic placement strategies.
\par
Firstly, for quasi-static detuning $\omega$, the drift-only evolution is given by (\ref{rhodrift}). Computing the random placement evolution (\ref{rhorand}) and subsequently the Haar-averaged squared trace distance, we obtain
\begin{align}
\int d\mu_2\, D^2
\propto
\,&\frac{1}{2}\left(\frac{\sin(\omega T)}{\omega T}-1\right)^2
-4\left(\frac{\sin(\omega T)}{\omega T}-1\right)\notag\\
&+\frac{4}{(\omega T)^2}\bigl(1-\cos(\omega T)\bigr)
-1-\cos(\omega T).
\end{align}
In the small-phase regime $\omega T \to 0$, this becomes
\begin{align}
\int d\mu_2\,D^2
&\propto
(\omega T)^2-\frac{1}{18}(\omega T)^4+O\!\bigl((\omega T)^6\bigr).
\end{align}
For deterministic placement at a fixed $t$, the corresponding Haar-averaged result is computed earlier in (\ref{Ddrift2}). The middle placement and front/back placement cases follow
\begin{align}
t=\frac{T}{2}:\
\int d\mu_2\, D^2
&\propto
8-2\left(\cos\frac{\omega T}{2}+1\right)^2\notag\\
&\to
(\omega T)^2-\frac{5}{96}(\omega T)^4+O\!\bigl((\omega T)^6\bigr),\\
t=0\ \text{or}\ T:\
\int d\mu_2\, D^2
&\propto
8-4\bigl(\cos(\omega T)+1\bigr)\notag\\
&\to
2(\omega T)^2-\frac{1}{6}(\omega T)^4+O\!\bigl((\omega T)^6\bigr).
\end{align}
Hence again, in the regime $\omega T\approx 0$, random placement behaves similarly to middle placement up to second order. And compared to front/back placement, it reduces the coefficient of the leading order term $(\omega T)^2$ by a factor of two.
This result constitutes the Haar-averaged analogue of the $|1\rangle$ pedagogical example: timing randomization suppresses a portion of the leading coherent contribution.
\\
\par
We next analyze random placement in the presence of amplitude damping and dephasing only, modeled by the combined quantum channel with Kraus operators given in (\ref{ADkraus}). Using (\ref{rhoAD}), the Haar-averaged squared trace distance for random placement is obtained as
\begin{align}
\int d\mu_2\, D^2
\propto&
3\bigl(F+f_2\bigr)^2
+3\bigl(1-f_1\bigr)^2+2\bigl(F+1\bigr)^2
+\bigl(1-e^{-T/T_2}\bigr)^2,
\end{align}
where
\begin{align}
f_1&=\frac{T_1}{T}\left(1-e^{-T/T_1}\right),\quad
f_2=\frac{T_2}{T}\left(1-e^{-T/T_2}\right),
\notag\\
F&=\frac{1}{T}\left(\frac{1}{T_1}-\frac{1}{T_2}\right)^{-1}
\left(e^{-T/T_1}-e^{-T/T_2}\right).
\end{align}
In the short-window limit $T\to 0$, we have
\begin{align}
\int d\mu_2\, D^2
&\propto
\frac{T^2}{2T_1^2T_2^2}\Bigl(3T_1^2+2T_1T_2+4T_2^2\Bigr)
+O(T^3).
\end{align}
For deterministic placements (midpoint and edges) (\ref{DT1T2}), the small $T$ expansions are given by
\begin{align}
t=\frac{T}{2}:\ \ \
&\frac{T^2}{2T_1^2T_2^2}\Bigl(3T_1^2+2T_1T_2+4T_2^2\Bigr)
+O(T^3),\\
t=0\ \text{or}\ T:\ \ \
&\frac{2T^2}{2T_1^2T_2^2}\Bigl(T_1^2+2T_2^2\Bigr)
+O(T^3).
\end{align}
Therefore, in the $T\to 0$ regime, random placement exhibits the same leading-order behavior as midpoint placement, while the front and back placements usually produce a larger error coefficient. This supports the practical interpretation of the random placement as a “smoothing” strategy: for sufficiently short idling intervals, the time-twirled channel matches the midpoint schedule at leading order, while also offering the potential benefit of suppressing timing-dependent coherent components when they are present, as seen clearly in the drift-only case.
\par
Finally, since the combined $T_1 T_2$+drift Haar-averaged expression (\ref{Dallnoise}) retains strong symmetry about $t=T/2$, it is natural to expect the same qualitative conclusion — random placement approximates midpoint behavior in the small-$T$ limit — to persist in the joint noise model. More broadly, since the mechanism underlying both midpoint and random placement relies on temporal symmetry and the redistribution of idle-time exposure, its qualitative effect is expected to extend beyond the specific noise models considered in this work.

\section{Conclusion}
In this paper, we investigated idling errors arising from amplitude damping, dephasing, and quasi-static frequency drift, and evaluated their impact using the trace distance. Building on this analysis, we proposed an error suppression approach that requires no additional control pulses, by adjusting the timing of H gates within available idle intervals. We studied two heuristic yet practical strategies — midpoint placement and shot-by-shot random placement — and theoretically showed that both can suppress decoherence and provide robust choices when the pre-H state is unknown. Hardware experiments and numerical simulations on single- and two-qubit circuits validated that these scheduling rules improve performance over edge placements across experimentally relevant idle durations, with random placement providing comparable benefits to midpoint placement in the short-idle regime.
\par
Our approach is complementary to dynamical decoupling. Whereas dynamical decoupling actively suppresses noise through the application of additional control pulses, our method suppresses idling errors by redistributing idle-time exposure without introducing any extra gate operations. Therefore, the two approaches target different operating regimes and resource constraints. A direct experimental comparison with dynamical decoupling is beyond the scope of this work, as it would require careful pulse-level calibration and introduce additional control-related errors, which might obscure the isolated effect of gate scheduling considered here. 
\par
Rather than proposing a new pulse-level control technique, this work identifies and formalizes a previously underexplored degree of freedom in quantum circuit execution: the temporal placement of operations within available idle intervals. We demonstrate that even simple symmetry-based scheduling strategies can provide robust suppression of idling-induced errors without additional hardware resources or control gate overhead. We therefore expect that integrating the proposed timing strategies with existing compilation, scheduling, and error-mitigation techniques will constitute a promising direction for future research.
\par
Other future works include extending the analysis to larger multi-qubit circuits and code-space-restricted state families (rather than Haar-random states) which will better quantify
the benefits for practical algorithms and QEC workloads, incorporating additional noise mechanisms such as residual ZZ, crosstalk, and leakage. 
\section*{Acknowledgment}
The authors would like to thank our colleagues Jun Fujisaki, Masatoshi Ishii, Mitsuki Katsuda and Tatsuya Amitani for valuable discussions and constructive feedback throughout this project, as well as for their helpful advice on the hardware experiments and numerical simulations, which helped improve this work.

\appendix

\section{Haar random initial state}\label{Haarrand}
\par
In many circuit-level analyses, the state of a qubit immediately before a Hadamard gate (or any gate of interest) is not known a priori: it depends on the preceding algorithmic context, intermediate entanglement, and earlier noise. To obtain a state-independent criterion, it is natural to average the trace distance over all possible single-qubit input states. In this work, we follow the standard approach of sampling initial states from the Haar measure \cite{Nielsen:2012yss}, i.e., treating the unknown pre-H state as a Haar-random pure state and computing the Haar-averaged trace distance between the noisy and ideal outputs.
\par
A Haar random single-qubit state (HRS) can be generated by applying a general rotation unitary $U\in \text{SU}(2)$ to a fixed reference state, typically $|0\rangle$. Using a Z-Y-Z Euler-angle parametrization, a general single-qubit unitary can be written as
\begin{align}
    U(\phi,\theta,\omega)
    &=
e^{-i\phi Z/2}\,e^{-i\theta Y/2}\,e^{-i\omega Z/2}\notag\\
&=
\begin{pmatrix}
e^{-i(\phi+\omega)/2}\cos(\theta/2) & -e^{-i(\phi-\omega)/2}\sin(\theta/2)\\
e^{i(\phi-\omega)/2}\sin(\theta/2) & e^{i(\phi+\omega)/2}\cos(\theta/2)
\end{pmatrix}.
\end{align}
The Haar measure corresponds to the parameter distribution
\begin{align}
\phi, \omega &\sim \mathrm{Uniform}(0,2\pi),
\notag\\
\theta &\sim \frac{1}{2}\sin\theta \; d\theta,
\quad \theta \in [0,\pi].
\end{align}
Applying this unitary to $|0\rangle$ produces the general pure state. Since global phase is physically irrelevant, the resulting density matrix depends only on $\theta$ and $\phi$. Explicitly, it is given by
\begin{align}
\rho_{\mathrm{HRS}}
=
\begin{pmatrix}
\cos^2\frac{\theta}{2}
&
e^{-i\phi}\cos\frac{\theta}{2}\sin\frac{\theta}{2}
\\
e^{i\phi}\cos\frac{\theta}{2}\sin\frac{\theta}{2}
&
\sin^2\frac{\theta}{2}
\end{pmatrix}.
\end{align}
Equivalently, in Bloch-vector form,
\begin{align}
    \rho_{\mathrm{HRS}}=\frac{1}{2}\Bigl(I+\mathbf{n}(\theta,\phi)\cdot\boldsymbol{\sigma}\Bigr),
\end{align}
where $\mathbf{n}(\theta,\phi)=
\bigl(\sin\theta\cos\phi,\ \sin\theta\sin\phi,\ \cos\theta\bigr)$ and $\boldsymbol{\sigma}=(X,Y,Z)$ is the vector of Pauli operators.
\par
Using $\rho_{\mathrm{HRS}}$ as the initial state, we compute the noisy final state (\ref{finalstate}) and evaluate its trace distance from the ideal state (\ref{idealstate}). To remove dependence on a particular (unknown) pre-H input state, we average this trace distance over the Haar measure
\begin{align}
    d\mu_2 = \sin\theta\, d\theta\, d\phi\, d\omega.
\end{align}
The Haar-averaged trace distance is then
\begin{align}
\overline{D}(t)
=
\frac{\displaystyle 1}
{\displaystyle \int d\mu_2}\int D\!\left(\rho(T,t),\,\rho'\right)\, d\mu_2.
\end{align}
Finally, the optimal timing $t^\star$ setting is chosen by minimizing the Haar-averaged distance:
\begin{align}
    t^\star = \arg\min_{t \in [0,T]} \ \overline{D}(t).
\end{align}
This procedure yields the gate timing that minimizes the average trace distance over all possible single-qubit input states, providing a robust timing criterion independent of prior state knowledge.

\bibliographystyle{unsrtnat}
\bibliography{references}

\end{document}